\documentstyle[aps,prl,epsfig,psfrag]{revtex}

\begin{document}
\draft
\title{Paramagnetic Breakdown of Superconductivity in Ultrasmall Metallic
  Grains} 
\author{Fabian Braun$^a$, Jan von Delft$^a$, D. C. Ralph$^b$ and 
M. Tinkham$^c$}
\address{$^a$Institut f\"ur Theoretische Festk\"orperphysik, Universit\"at
  Karlsruhe, 76128 Karlsruhe, Germany\\
  $^b$Laboratory of Atomic and Solid State Physics, 
  Cornell University, Ithaca, NY 14853.\\
  $^c$Department of Physics and Division of Engineering and Applied Science,
  Harvard University, Cambridge, MA 02138}

\twocolumn[\hsize\textwidth\columnwidth\hsize\csname%
@twocolumnfalse\endcsname%
\date{June 10, 1997}%
\maketitle%
\begin{abstract}
  We study the magnetic-field-induced breakdown of superconductivity in
  nm-scale metal grains having a mean electron level spacing $d \simeq
  \tilde\Delta$ (bulk gap). Using a generalized variational BCS approach that
  yields good qualitative agreement with measured spectra, we argue that Pauli
  paramagnetism dominates orbital diamagnetism, as in the case of thin films
  in a parallel magnetic field.  However, the first-order transition observed
  for the latter can be made continuous by finite size effects. The
  mean-{}\linebreak{}field procedure of describing the system by a single
  pairing parameter $\Delta$ breaks down for $d \simeq \tilde\Delta$.
\end{abstract}
\vskip0.0pc]    

When a system of (correlated) electrons is sufficiently small, the electronic
spectrum becomes discrete.  This allows one to study the nature of electron
correlations in unprecedented detail by analyzing the details of the spectrum.
It has recently become possible to measure such discrete spectra directly by
studying electron transport through nm-scale metallic grains (radius
$r\approx5$nm), for which the mean spacing $d=1/{\cal N}(\varepsilon_F)$ is
$\simeq 0.1- 1$~meV \cite{RBT95-96,RBT}.  For Al grains the effects on the
spectrum of spin-orbit interactions \cite{RBT95-96}, non-equilibrium
excitations \cite{agam} and superconductivity \cite{RBT95-96,jan,zaikin} have
been investigated.

Studying the latter is particularly interesting in grains with $d \simeq
\tilde \Delta$ (bulk gap), near the lower size limit \cite{anderson} of
observable superconductivity. The number of free-electron states that
pair-correlate (those within $\tilde \Delta $ of $\varepsilon_F$) is then of
order one. Thus, even in grains in which a gap can still be observed
\cite{RBT95-96}, pairing correlations are expected to become so weak that they
might be destroyed by the presence of a single unpaired electron \cite{jan}.
A direct way to probe this is to turn on a magnetic field, whose Zeeman energy
favors paramagnetic states with non-zero total spin.

In this Letter, we develop a theory for the paramagnetic breakdown of
superconductivity in nm-scale grains. We exploit analogies to thin films in a
parallel magnetic field \cite{Meservey}, but explicitly take account of the
discreteness of the spectrum.  To calculate the eigenenergies $E_n$ of the
grain's lowest-lying eigenstates $|n\rangle$, we adopt a generalized
variational BCS approach that goes beyond standard mean-field theory by using
a different pairing parameter $\Delta_n$ for each $|n \rangle$.  Using the
$E_n$ to reconstruct the tunneling spectra, we find qualitative agreement with
measured spectra \cite{RBT95-96}, and show that the $H$-induced first-order
transition to the paramagnetic normal state observed for thin films can be
softened in ultrasmall grains.

{\it Experimental Results.---} Our goal is to understand in detail the
$H$-dependence of the measured discrete tunneling spectrum (see Fig.~3 of
\cite{RBT}) of an ultrasmall Al grain, coupled via tunnel barriers to one gate
and two lead electrodes to form a nm-scale transistor.  Each line in the
spectrum corresponds to the $H$-dependent energy $E^{N}_{n_f}-E^{N \pm
  1}_{n_i}+(E_C^N - E_C^{N\pm1})$ \vspace{-.2mm} needed for some rate-limiting
electron tunneling process $|n_i\rangle_{N\pm1} \to |n_f\rangle_N$ off or onto
the grain, where $|n\rangle_N$ is an eigenstate (with eigenenergy $E^N_{n} +
E_C^N$) of the $N$-electron island with charging energy $E_C^N$.  Since the
change in charging energy $\delta E_C$ depends on the adjustable gate voltage
$V_g$, so does the odd-even ground state energy difference $(E_G^{N+1} -
E_G^N)$ (so that the BCS gap for the ground state of an odd grain can not be
measured absolutely). Therefore only the energy differences $E^{N}_{n_f} -
E^{N}_{{n_f}'}$\vspace{-.3mm} between same-$N$ final states of transitions
with the same initial $|n_i\rangle_{N\pm1}$, i.e. the spacing {\em between}
lines of a given spectrum, are physically significant. They give the grain's
fixed-$N$ eigenspectrum.  By appropriately adjusting $V_g$, both even and odd
spectra ($N\!=\!2m+p$, with $p=(0,1)$ for $(e,o)$ parity) were measured, and
non-equilibrium effects \cite{RBT,agam} minimized.

The presence (absence) of a clear spectroscopic gap $2 \Omega \! > \! d$
between the lowest two lines of the odd-to-even (even-to-odd) measured
spectra (Fig.~3(a,b) of \cite{RBT}) reveals the presence of pairing
correlations \cite{RBT95-96,jan}: in even grains, all excited states involve at
least two BCS quasi-particles and hence lie significantly above the ground
state, whereas odd grains {\em always}\/ have at least one quasi-particle and
excitations need not overcome an extra gap.

\fussy
{\em Pauli paramagnetism.---} The measured levels' approximately linear
behavior with $H$ can be attributed to the electrons' Zeeman energies $\pm h
\equiv \pm {1 \over 2} \mu_B g H $; indeed, the differences between measured
slopes of up- and down-moving lines correspond to $g$-factors between $1.95$
and~$2$.  (Deviations from $g = 2$ probably result from spin-orbit scattering,
known to be small but non-zero in thin Al films \cite{Meservey}, but neglected
below.)  Thus, the $H$-dependence of the spectroscopic gap $\Omega (H)$ is
almost entirely of Zeeman origin.  Note that $\Omega (H)$ must be
distinguished from the BCS pairing parameter $\Delta (H)$. In contrast to bulk
samples, in ultrasmall grains the suppression of $\Delta(H)$ through orbital
diamagnetism 
is very weak, just as in thin films in parallel fields \cite{Meservey}: The
flux through the grain (whose radius $5$nm $\ll$ the penetration length of
50nm) is of order $5\%$ of a flux quantum $\phi_0$ at $H=7$T, i.e. too small
to significantly affect the orbital motion of electrons between reflections
off the grain boundaries.  Slight deviations from $H$-linearity observed
in some larger grains \cite{RBT95-96} probably reflect the onset of
orbital diamagnetism (giving corrections to eigenenergies of order
${\stackrel{\scriptscriptstyle<}{\scriptscriptstyle\sim}}\, \hbar v_F r^3
(H/\phi_o)^2$,~\cite{Bahcall}). But for the spectra of interest here, they are
much smaller than Zeeman effects and hence will be neglected.

\sloppy
Now, Clogston and Chandrasekhar (CC) \cite{clogston} argued that in the
absence of orbital diamagnetism, superconductivity will be destroyed by Pauli
paramagnetism: Let $|s\rangle$ denote the ground state of the spin-$s$ sector
($s=J + p/2$ with $J$ an integer) of the $(2m+p)$-electron Hilbert space
\cite{muehlschlegel}, with exact eigenenergy $E_s(h,d) = E_s(0,d) - 2 s h$
(``spin'' simply means $\sum s_z$ with $s_z = \pm {1\over2}$). CC pointed out
that a ground state transition will occur from $|p/2\rangle$ to some {\em
  normal}\/ state $|\bar s \rangle$ when $E_{p/2}(\bar h,d) = E_{\bar s}(\bar
h,d)$ at some sufficiently large field $\bar h$.  For $d\ll \tilde \Delta$ (as
in thin films), $E_{p/2}(0,d) = - \tilde \Delta^2 / 2d$ and $E_{\bar s}(0,d) =
(\bar s^2 -p/4) d$, with $\bar s \simeq \bar h/d$ [to ensure $\partial_{\bar
  s} E_{\bar s}(h,d) = 0$].  Thus, CC predicted a first-order transition at a
critical field $\bar h_{CC}=\tilde \Delta/\sqrt2$, the new spin $\bar
s=\tilde\Delta/\sqrt2d$ being macroscopically large.  In tunneling
measurements \cite{Meservey} into thin (5nm) Al films ($\tilde \Delta =
0.38$meV and $H_{\rm CC} = 4.7$T) this first-order transition was observed as
a {\em jump} in the tunneling threshold from $\Delta - h_{\rm CC}$ to 0 at
$\bar h_{\rm CC}$.  In contrast, the measured energy levels for ultrasmall
grains evolve {\em continuously}\/ with $h$, showing kinks but no jumps. We
suggest that this reflects a ``softening'' of the transition that occurs when
$d \!  \simeq \!  \tilde \Delta$, for which $\bar s$ should become of order
one. We shall show this explicitly by performing model calculations of the
energies $E_{\bar s} (h,d)$ of BCS-like pair-correlated variational states $|s
\rangle$.

{\it The model.---} We adopt the reduced BCS Hamiltonian used in
Ref.~\cite{jan} with an additional Zeeman term:
\begin{equation} 
\label{hamilton}
  H = \sum_{j\sigma}
  (\varepsilon_j+\sigma h)   c^\dagger_{j\sigma}c_{j\sigma}
    - \lambda d \sum_{j,j'}
    c^\dagger_{j+}c^\dagger_{j-}
      c_{j'-}c_{j'+} \, .
\end{equation}
$c_{j \pm}^\dagger$ create free time-reversed states $|j,\pm\rangle$, whose
energies $\varepsilon_j = j d + \varepsilon_0 - \mu$, measured relative to the
chemical potential $\mu$, are taken uniformly spaced for simplicity (though
this is not essential \cite{ambegaokar}).  $j=0$ labels the first level whose
occupation in the $T=0$ Fermi sea is not 2 but $p$, and $|F \rangle =
\prod_{j= -m}^{-1} c_{j +}^\dagger c_{j -}^\dagger |\mbox{Vac}\rangle $ is the
even Fermi sea.

{\it Variational approach.---} Since in the experiments $T = 50$mK${}\ll d ,
\tilde \Delta$, we set $T=0$.  For the spin-$s$ ground state $|s \rangle$, we
make a generalized BCS-like variational Ansatz \cite{Soloviev}, which
pair-correlates all time-reversed states, except for $2s$ unpaired spin-up
electrons placed as close as possible to the Fermi surface, to minimize the
kinetic energy cost of having more spin ups than downs:
\begin{eqnarray}
\label{ansatzn}
    |s\rangle = \prod_{j=-J}^{J+p-1} c^\dagger_{j+} 
                \prod'_j (u^{(s)}_j + v^{(s)}_j
       c^\dagger_{j+}c^\dagger_{j-})\,|\mbox{Vac}\rangle \; .
\end{eqnarray}                                               
The prime over products (and over sums below) indicates exclusion of the
singly occupied states $j=-J$ to $J+p-1$ (for which $u^{(s)},v^{(s)}$ are not
defined).  Since $\langle s | s' \rangle = \delta_{ss'}$, the variational
parameters $v_j^{(s)}$ and $u_j^{(s)}$ must be found {\em independently\/} for
each $s$ (hence the superscript $s$), by minimizing the variational
``eigenenergies''
\begin{eqnarray}
\label{Es}
\lefteqn{
   {\cal E}_s (h,d) \: \equiv \: \langle s | H | s \rangle  \: = \: 
-2hs + \sum_{j=-J}^{J+p-1}\varepsilon_j } \\ \nonumber
 & & + 2\sum'_j\varepsilon_j (v^{(s)}_j)^2
   -  \lambda d\Big(\sum'_{j}u^{(s)}_jv^{(s)}_j\Big)^2 + \lambda d
    \sum'_j (v^{(s)}_j)^4,
\end{eqnarray}
which we use to approximate the exact eigenenergies $E_s(n,d)$.  The
$v^4$-term, not extensive and hence neglected in the bulk case, is
non-negligible, but not dominant either.  Solving $\partial {\cal E}_s/
\partial v^{(s)}_j \!=\! 0$ and $u^2 \!+\! v^2\! =\! 1$ simultaneously yields
$(v_j^{(s)})^2 \!=$ $(1 \!-\! \bar \varepsilon_j / [\bar \varepsilon_j^2 \!+\!
\Delta_s^2]^{1/2})/2 $, with $\Delta_s$ determined by the generalized gap
equation $\Delta_s \!=\! \lambda d \sum'_j u_j^{(s)} v_j^{(s)}$, and $\bar
\varepsilon_j \!=\! \varepsilon_j \!-\!  \lambda d (v_j^{(s)})^2$.  (To
simplify our calculations, we used $\bar \varepsilon_j \!=\! \varepsilon_j$;
this changes ${\cal E}_s$, which is stationary under small changes in
$v_j^{(s)}$, only to order $\lambda^2$.  Then $\langle \hat N \rangle \!=\!
2m+p$ fixes the chemical potential to $\mu=\varepsilon_0 \!-\! \delta_{p,0}\,
d/2 $.)

\begin{figure}[t]
  \begin{center}
    \leavevmode
    \psfrag{0}[r][r]{\footnotesize0}
    \psfrag{1}[r][r]{\footnotesize1}
    \psfrag{1.5}[t][t]{\footnotesize1.5}
    \psfrag{2}[t][t]{\footnotesize2}
    \psfrag{-2}[t][t]{\footnotesize-2}
    \psfrag{3}[t][t]{\footnotesize3}
    \psfrag{3.5}[t][t]{\footnotesize3.5}
    \psfrag{0.5}[t][t]{\footnotesize0.5}
    \psfrag{d}[l][l]{$d/\tilde{\Delta}$}
    \psfrag{dt}[][][1][90]{$\Delta_s/\tilde\Delta$}
    \psfrag{en}[][][1][90]{${\cal E}_{s}/\tilde\Delta$}
    \psfrag{n0}[][]{\footnotesize$\Delta_0$}
    \psfrag{n1}[][]{\footnotesize$\Delta_{1/2}$}
    \psfrag{n2}[][]{\footnotesize$\Delta_1$}
    \psfrag{s=0}[][]{\footnotesize$s=0$}
    \psfrag{12}[][]{\footnotesize$1$}
    \psfrag{s1}[][]{\footnotesize$\!\frac32$}
    \psfrag{32}[][]{\footnotesize$2$}
    \epsfig{width=0.9\linewidth,file=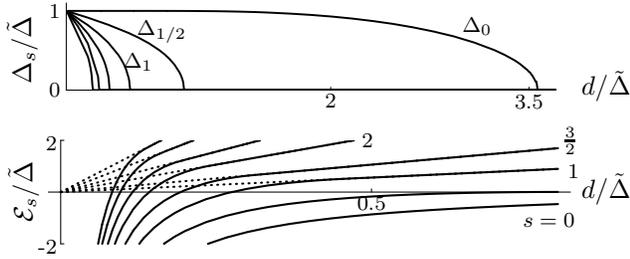}\vspace*{2mm}
    \caption{(a) Pairing parameters $\Delta_s(d)$ 
      for some spin-$s$ states $|s\rangle$, as a function of level spacing $d$
      ($\Delta_{0,1/2} = \Delta_{e,o}$ of \protect\cite{jan}).  (b) The
      variational energies $\langle s| H|s \rangle - \langle F|H|F\rangle$,
      plotted as functions of $d$ at magnetic field $h=0$ (solid lines,
      $\approx -\tilde\Delta^2/2d$ as $d\to0$) smoothly approach the energies
      ${}_0\langle s| H|s \rangle_0$ $ - \langle F|H|F\rangle = (s^2 -p/4) d$
      (dotted lines) of the uncorrelated states $|s\rangle_0$, since $\Delta_s
      (d) \to 0$ with increasing $d$.}
    \label{fig:2}
  \end{center}
\vspace{-7mm}
\end{figure}

{\em The pairing parameter $\Delta_s$.---} In the variational approach,
$\Delta_s$ is merely an auxiliary quantity in terms of which the $ v_j^{(s)}$
and hence ${\cal E}_s$ are parameterized, and certainly not directly
measurable.  It does serve as a measure of the pairing correlations present in
$|s\rangle$, though: if $\Delta_s = 0$, $|s \rangle$ reduces to the
paramagnetic spin-$s$ ground state $|s\rangle_0 =
\prod_{j=0}^{J+p-1}c_{j+}^\dagger \prod_{j=-J}^{-1} c_{j-} |F\rangle$, and the
``correlation energy'' $\langle s| H|s \rangle - {}_0\langle s| H|s \rangle_0$
vanishes (see Fig.~\ref{fig:2}b).  The gap equation for $\Delta_s(d)$ is
$h$-independent (since the $h$-dependence of ${\cal E}_s$ is so trivial), and
differs from the standard bulk $T=0$ case (for which $s=p/2$,
$d\ll\tilde\Delta$) through both the discreteness and the $s$-dependent
restriction on the sum, which respectively cause the $d$- and $s$-dependence
of $\Delta_s(d)$.  Its numerical solution (see Fig.~\ref{fig:2}a) shows that
$\Delta_s(d)$ decreases to zero as $d$ is increased, because the kinetic
energy cost of pairing correlations, which shift electron occupation
probability from below to above $\varepsilon_F$, grows with increasing $d$
\cite{jan}. Moreover, $\Delta_s(d)$ {\em decreases rapidly with increasing}\/
$s$ at fixed $d$ (reaching zero roughly at $d= \tilde \Delta/2s$, as can be
shown analytically).  This $s$-dependence of $\Delta_s$, a generalization of
the even-odd effect (namely $\Delta_{1/2} < \Delta_0$) found in~\cite{jan}, is
so strong because for $d \simeq \tilde \Delta$ there are only a few
$\varepsilon_j$'s within $\tilde \Delta$ of $\varepsilon_F$ (where pairing
correlations are strongest), so that increasing $s$ and hence the number of
unpaired electrons in this regime {\em dramatically reduces the strength of
  pairing correlations}. Evidently, for $d \simeq \tilde \Delta$ the standard
mean-field description of superconductivity in terms of only a {\em single\/}
pairing parameter $\Delta$ is no longer sufficient.

Conceivably the ${\cal E}_s$, which are upper bounds on the exact spin-$s$
ground state energies $E_s$, can be lowered by using better variational
wavefunctions that sample a larger portion of the spin-$s$ Hilbert space,
i.e.\ by ``\mbox{including} fluctuations'' about the variational state
$|s\rangle$.  \mbox{But in} the present context this would hardly be
worthwhile, since the $E_s$ also depend quite sensitively on the unknown input
energies $\{ \varepsilon_j\}$ \cite{ambegaokar}, and here we merely seek a
qualitative understanding of the measured tunneling spectra.

\begin{figure}[t]
  \begin{center}
    \leavevmode
    \psfrag{0}[r][r]{\footnotesize0}
    \psfrag{1-}[r][r]{\footnotesize$1-\frac{1}{\sqrt{2}}$}
    \psfrag{1+}[r][r]{\footnotesize$\frac{1}{\sqrt{2}}$}
    \psfrag{1}[r][r]{\footnotesize1}
    \psfrag{0.4}[t][t]{\footnotesize0.4}
    \psfrag{0.8}[t][t]{\footnotesize0.8}
    \psfrag{0.2}[t][t]{\footnotesize0.2}
    \psfrag{0.6}[t][t]{\footnotesize0.6}
    \psfrag{d}[r][r]{$d/\tilde{\Delta}$}
    \psfrag{dt}[][][1][90]{$h_{s',s}/\tilde\Delta$}
    \psfrag{r1}[][][1][90]{\footnotesize I}
    \psfrag{r2}[][][1][90]{\footnotesize II}
    \psfrag{r3}[][][1][90]{\footnotesize III}
    \psfrag{r5}[][][1][90]{\footnotesize V}
    \psfrag{0-2}[][]{\footnotesize $(0,1)$}
    \psfrag{0-4}[][]{\footnotesize $(0,2)$}
    \psfrag{1-3}[][]{\footnotesize $(\frac12,\frac32)$}
    \psfrag{2-4}[][]{\footnotesize $(1,2)$}
    \psfrag{3-5}[][]{\footnotesize $(\frac32,\frac52)$}
    \epsfig{width=0.9\linewidth,file=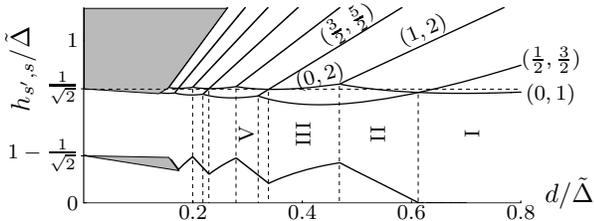}\vspace*{2mm}
    \caption{$d$-dependence of the critical fields $h_{s,s'}$, at which $h$
      induces, at {\em fixed} electron number $N$, a ground state transition
      from $|s\rangle$ to $|s'\rangle$, labeled by $(s,s')$ (in contrast we
      label the $N$-{\em changing} tunneling transitions in Fig.~\ref{fig:3}
      by $|n\rangle\to|n'\rangle$). The lower curve gives the jump at $\bar
      h_{0 ,\bar s}(d)$ predicted for the lowest line of the $e\to o$
      tunneling spectra of Fig.~\ref{fig:3}.}
    \label{fig:hcrit}
  \end{center}
\vspace{-4mm}
\end{figure}

{\it Critical Fields.---} Having obtained $\Delta_s (d)$, we find the energies
${\cal E}_s(h, d)$ numerically from Eq.~(\ref{Es}).  For two same-parity spins
$s'=s+J$, ${\cal E}_{s'}$ drops below ${\cal E}_{s}$ at the field:
\begin{equation}
  h_{s,s'} (d) = [{\cal E}_{s'} (0,d)-{\cal E}_{s}(0,d)] / (2s'-2s).
\end{equation}
Fig.~\ref{fig:hcrit} shows several $ h_{s,s'} $ as functions of $d$.  For
given $p$, let their lower envelope be denoted by $\bar h_{p/2, \bar s} (d) =
\mbox{min}[h_{p/2,s'}(d)]$.  This gives the ``critical field'' at which, if
$h$ is increased from 0 at fixed $d$ and $p$, the {\em first\/} change of
ground state occurs from $|p/2 \rangle$ to a new ground state $|\bar s
(d)\rangle$, whose spin depends on $d$.  Numerically we find $\bar h_{p/2,
  \bar s} (d \to 0) = \tilde \Delta / \sqrt 2$, which is CC's bulk result.  We
also find that {\em for any $d$}, $|\bar s (d) \rangle$ always has
$\Delta_{\bar s} (d) = 0$.  Thus, the first ground state transition is always
directly into a normal paramagnetic state with no pairing correlations. As
anticipated above, the change in spin, $\bar s-p/2$, macroscopically large at
small $d/\tilde \Delta$, {\em decreases with increasing $d$}. In this sense
the first-order transition observed in thin films is ``softened'' for
ultrasmall grains.  $\bar s-p/2$\vspace{-.3mm} reaches unity when the
correlation energy becomes smaller than the Zeeman energy gained by flipping a
single spin: for $p=0$ (or 1) we find $\bar s = 1$ (or 3/2) when $d/ \tilde
\Delta > 0.47$ (or 0.32).  This regime displays ``minimal superconductivity'':
correlations are strong enough to cause a measurable gap, yet so weak that
breaking a {\em single\/} pair destroys them.

{\it Tunneling Spectra.---} In analogy to $|s\rangle$, one can also define
excited states $|s,n\rangle$ in the spin-$s$ sector of Hilbert space by
placing the unpaired spins further away from the Fermi level, and
variationally calculate their energies ${\cal E}_{s,n} (h,d)$ (writing ${\cal
  E}_s = {\cal E}_{s,0}$). It is easy now to reconstruct the expected
tunneling spectra as function of $h$ at fixed $d$, by finding the energy cost
for all $|s_i, n_i\rangle_{N\pm1} \to |s_f,n_f\rangle_N $ single-electron
tunneling transitions. Since these satisfy the selection rule $s_f - s_i = \pm
1/2$, only slopes of $\pm 1$ can occur. Neglecting non-equilibrium effects
\cite{RBT,agam}, we always take the ground state of a given spin-$s$ sector as
the initial state and denote it by $| s_{i} (h,d), 0\rangle$. The appropriate
$s_i(h,d)$ must be determined from Fig.~\ref{fig:hcrit}.

\begin{figure}[t]
  \begin{center}
    \leavevmode
    \psfrag{0}[t][t]{\footnotesize0}
    \psfrag{0.4}[t][t]{\footnotesize0.4}
    \psfrag{0.8}[t][t]{\footnotesize0.8}
    \psfrag{1.2}[t][t]{\footnotesize1.2}
    \psfrag{o}[l][l]{\footnotesize0}
    \psfrag{i}[l][l]{\footnotesize1}
    \psfrag{j}[lt][lt]{\footnotesize2}
    \psfrag{01}[][]{\footnotesize $0\!\to\!\frac12$}
    \psfrag{10}[][]{\footnotesize $\frac12\!\to\!0$}
    \psfrag{12}[][]{\footnotesize $\frac12\!\to\!1$}
    \psfrag{23}[][]{\footnotesize $1\!\to\!\frac32$}
    \psfrag{32}[][]{\footnotesize $\frac32\!\to\!1$}
    \psfrag{45}[][]{\footnotesize $2\!\to\!\frac52$}
    \psfrag{a}[][]{\footnotesize a)}
    \psfrag{b}[][]{\footnotesize b)}
    \psfrag{c}[][]{\footnotesize c)}
    \psfrag{d}[][]{\footnotesize d)}
    \psfrag{Magnetic Field}[][]{Magnetic Field $h/\tilde\Delta$}
    \psfrag{En}[][][1][90]{Energy $({\cal E}^{N}_{n} - {\cal
  E}^{N\pm1}_{n'})/\tilde\Delta$}
    \psfrag{oe,d=0.67}[b][b]{\parbox{16mm}{$o\to e$\\$d=0.67\tilde\Delta$}}
    \psfrag{eo,d=0.67}[b][b]{\parbox{16mm}{$e\to o$\\$d=0.67\tilde\Delta$}}
    \psfrag{oe,d=0.34}[b][b]{\parbox{16mm}{$o\to e$\\$d=0.34\tilde\Delta$}}
    \psfrag{eo,d=0.34}[b][b]{\parbox{16mm}{$e\to o$\\$d=0.34\tilde\Delta$}}
    \psfrag{h20}[r][r]{\footnotesize $h_{0,1}$}
    \psfrag{h42}[l][l]{\footnotesize $h_{1,2}$}
    \psfrag{h31}[l][l]{\footnotesize $h_{\frac12,\frac32}$}
    \psfrag{i31}[r][r]{\footnotesize $h_{\frac12,\frac32}$}
    \psfrag{h40}[l][l]{\footnotesize $h_{0,2}$}
    \epsfig{width=0.9\linewidth,file=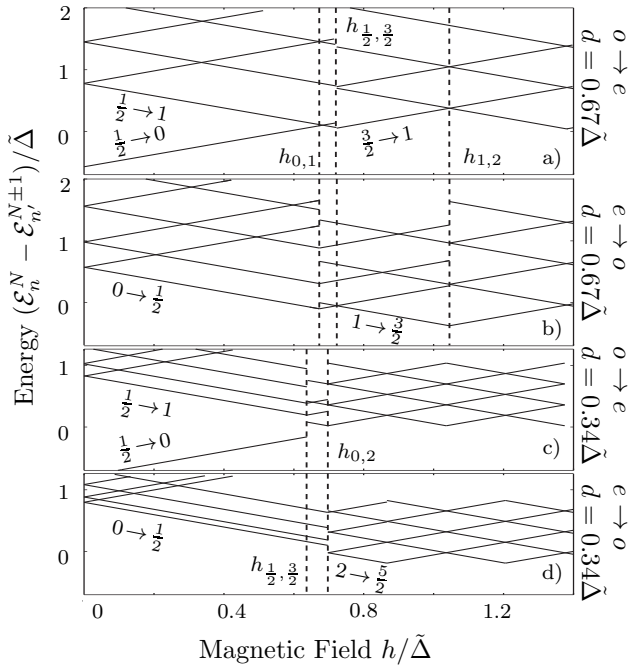}\vspace*{2mm}
    \caption{Tunneling spectra predicted for an ultrasmall superconducting
      grain for $d/\tilde\Delta=0.67$ and $0.34$. Some lines are labelled by
      the tunneling-induced change $s \to s'$ in the grain's spin.  For
      clarity, not all higher lines are shown.  Dashed lines indicate some of
      the critical fields $h_{s,s'}$.}
    \label{fig:3} \label{fig:1}
  \end{center}
  \vspace{-4mm}
\end{figure}

Whenever $h$ passes through one of the critical fields $h_{ s_i, s_i'}$, the
current ground state of the $N\pm 1$ electron Hilbert space changes to $| s_i'
\rangle $.  This produces a discontinuity in the lowest line of the tunneling
spectrum if $s_f - s_i'$ now violates the selection rule, or else a kink if
only its sign changes relative to $s_f - s_i$.  According to
Fig.~\ref{fig:hcrit}, depending on $d$ one can distinguish different regimes
I, II, III, \ldots, in each of which the various $ s_i$ to $ s'_i$ ground
state changes occur in a different order, leading to different magnitudes and
positions of jumps in the tunneling spectra.  In regime~I, where the order of
occurrence of ground state changes with increasing $h$ is $(0,1),
(\frac12,\frac32), (1,2), (\frac32,\frac52),\ldots$, there are no
discontinuities in the evolution of the lowest line [see
Fig.~\ref{fig:1}(a,b)].  E.g.\ for the $e\to o$ spectrum, the lowest $|0
\rangle \to |1/2\rangle$ line changes {\em continuously\/} to $|1\rangle \to
|1/2\rangle$ at $h_{0,1}$, since $s_f - s'_i = -1/2$.  However, in all other
regimes, $ \bar s > 1$ for the first ground state transition at $h_{0, \bar
  s}$, implying a {\em jump\/} in all $e\to o$ lines.  The jump's magnitude
for the lowest $e\to o$ line, shown as function of $d$ by the lower line in
Fig.~\ref{fig:hcrit}, starts at $d=0$ from the CC value $\tilde \Delta (1 -
1/\sqrt2)$ measured for thin Al films \cite{Meservey}, and decreases to 0
(non-monotonically, due to the discrete spectrum), again illustrating the
softening of the transition.

The absence of observable jumps in the measured lowest lines can be explained
by assuming the grain to lie in the ``minimal superconductivity'' regime I
\cite{footnote}. Indeed, the overall evolution of the lowest lines of Fig.~3
of \cite{RBT} qualitatively agrees quite well with those predicted for regime
I [Fig.~\ref{fig:1}({a,b})], in which the correlation energy is non-zero only
for $s = 0, 1/2$. The prediction of $h_{0,1} \simeq 0.95 h_{\rm CC}$ is
compatible with the experimentally observed value of $H_{0,1} = 4$T, which is
about $85\%$ of $H_{\rm CC} = 4.7$T for thin films \cite{Meservey}. The jumps
in {\em higher\/} lines (e.g. in Fig.~\ref{fig:3}(b) at $h_{1,2}$) are due to
correlations left in exited states $|s, n\rangle$ ($\Delta_{s,n}>\Delta_{s,0}$
since the unpaired electrons are further away from $\varepsilon_F$).
Experimentally, these jumps have not been observed.  This may be because
up-moving resonances lose amplitude and are difficult to follow with
increasing $h$ \cite{RBT}, or because the widths of the excited resonances
($\approx 0.13\tilde\Delta$) limit energy resolution \cite{agam}. More than
qualitative agreement between theory and experiment can not be expected, for
we assumed constant level spacing, neglected non-equilibrium effects, and the
tunneling matrix elements are unknown.

Our theory predicts that for somewhat larger grains (in regimes II or higher)
the tunneling spectra should show jumps even in the lowest line.  It remains
to be investigated, though, whether orbital diamagnetic effects, which rapidly
increase with increasing grain size ($\sim r^3$), would not smooth out such
jumps.

{\em Non-time-reversed pairing.---} By using the {\em reduced}\/ BCS
Hamiltonian in Eq.~(\ref{hamilton}), we neglected interaction terms $-d
\sum_{iji'j'} \lambda (i,j,i',j') c^\dagger_{i+}c^\dagger_{j-} c_{i'-}c_{j'+}
$ between non-time-reversed pairs $ c^\dagger_{i+}c^\dagger_{j-}$, following
Anderson's argument \cite{anderson} that for a short-ranged interaction, the
matrix elements involving time-reversed states $ c^\dagger_{j+}c^\dagger_{j-}$
are much larger than all others, since their orbital wave-functions interfere
constructively \cite{altshuler}. Interestingly, the experimental results
provide strikingly direct support for the correctness of purely time-reversed
pairing at $h=0$: if the $\lambda (j+\delta j,j, j'+ \delta j, j')$ were all
roughly equal to $\lambda$ for a finite range of $\delta j$, (instead of being
negligible for $\delta j \neq 0$, as assumed in $H_{\rm red}$), then for $2s <
\delta j$ one could construct spin-$s$ states with manifestly lower energy
than $|s\rangle$,
by pair-correlating {\em non\/}-time-reversed states, with the $2s$
uncorrelated electrons at the band's {\em bottom}\/, where 
having them uncorrelated costs hardly
any correlation energy:
\begin{eqnarray*}
    |s\rangle' = \prod_{j=-m}^{-m+2s-1} c^\dagger_{j+} 
                \prod_{j= -m}^{\infty} (u^{(s)}_j + v^{(s)}_j
       c^\dagger_{(j+2s)+}c^\dagger_{j-})\,|\mbox{Vac}\rangle \; .
\end{eqnarray*}
However, since it then costs
no correlation energy to increase $s$ (as can be checked explicitly), there
would be no large threshold for $h$ to induce ground state changes, and the
change $|p/2 \rangle$ to $|p/2 +1\rangle$ would occur at roughly $h \simeq d$
(as in a normal paramagnet), contradicting the experimental finding that the
first kinks only occur after a sizeable threshold $\bar h_{p/2,\bar s} \simeq
\tilde \Delta/ \sqrt{2}$.

{\it Conclusions.---} The dominant mechanism for destroying pairing
correlations in ultrasmall grains is Pauli paramagnetism.  Calculating the
variational eigenenergies of the lowest-lying eigenstates by a generalized
variational \linebreak BCS method, we have shown that decreasing grain size
softens the first-order transition observed for thin films by reducing the
number of spins flipped from being ma\-cros\-copically large to being of order
one for $d \simeq \tilde \Delta$. Our approach qualitatively reproduces the
measured tun\-nel\-ing thresholds, explaining why they do not show jumps.

We would like to thank V. Ambegaokar, S. Bahcall, D. Golubev, B. Janko, A.
Rosch, G.  Sch\"on, R. Smith and A. Zaikin for discussions. This research was
supported by ``SFB~195'' of the Deutsche For\-schungs\-ge\-mein\-schaft, NSF
Grant No.\ DMR-92-07956, ONR Grant No.\ N00014-96-1-0108, JSEP Grant No.\ 
N00014-89-J-1023, NSF MRSEC Grant No.\ DMR-9632275 and the AP Sloan
Foundation.\vspace{-4mm}

\end{document}